\documentclass[conference,a4paper]{IEEEtran}
\IEEEoverridecommandlockouts

\usepackage[T1]{fontenc}
\usepackage[utf8]{inputenc}
\usepackage{cite}
\usepackage{amsmath,amssymb}
\usepackage{graphicx}
\usepackage{booktabs}
\usepackage{url}
\usepackage{xspace}   

\newcommand{\method}{TAHA\xspace}
\newcommand{\arbiter}{VLM-as-Arbiter\xspace}

\begin{document}

\title{Cost-Aware Vision--Language Model Arbitration\\
       for Fabric Structure Recognition:\\
       A Deployable Multi-Agent System}

\author{%
  \IEEEauthorblockN{Chenwei~Wang\IEEEauthorrefmark{1},
                    Haochen~Li\IEEEauthorrefmark{1}\textsuperscript{,}\IEEEauthorrefmark{2},
                    Shuk~Ching~Tang\IEEEauthorrefmark{3},
                    Misbah~Iqbal\IEEEauthorrefmark{3},
                    Carman~K.~M.~Lee\IEEEauthorrefmark{3},
                    Elif~Ozden-Yenigun\IEEEauthorrefmark{1}\textsuperscript{,}\IEEEauthorrefmark{2}%
    \thanks{Accepted for publication in the Proceedings of the IEEE
      International Conference on Industrial Engineering and Engineering
      Management (IEEM) 2026, Singapore. This is the authors' version;
      the definitive version will appear in IEEE \emph{Xplore}.
      \copyright~2026 IEEE. Personal use of this material is permitted.
      Permission from IEEE must be obtained for all other uses.}%
    \thanks{Corresponding author: Elif Ozden-Yenigun
      (\texttt{elif.ozden-yenigun@rca.ac.uk}).}}%
  \IEEEauthorblockA{\IEEEauthorrefmark{1}Laboratory for Artificial
                    Intelligence in Design (AiDLab), PolyU/RCA,
                    Hong Kong SAR / London, UK\\
                    \IEEEauthorrefmark{2}Royal College of Art, London, UK\\
                    \IEEEauthorrefmark{3}Department of Industrial and Systems
                    Engineering, The Hong Kong Polytechnic University,
                    Hong Kong SAR}%
}

\maketitle

\begin{abstract}
Recognizing a fabric's structure is a prerequisite for translating
textile-specific material information into structured digital form for
downstream supply-chain systems. Pure CNN classifiers are cost-efficient
but fail on visually ambiguous categories; vision--language models (VLMs)
generalize more broadly but cost much more per image and are unstable on
specialist domains. We present a multi-agent system in which a CNN cascade
handles the easy majority and a VLM is invoked only as a selective arbiter,
constrained to a top-3 taxonomy-consistent choice. The fabric taxonomy
performs as a constraint for the whole recognition process to increase the
accuracy and reduce the VLM calls. Meanwhile, the CNN cascade is distilled
to a small parameter size to reduce the inference time and meet the needs of
practical deployment. On a newly curated 14-class benchmark, a flat
ConvNeXt-Tiny baseline reaches $90.45\,\%$ top-1 and $76.9\,\%$ on the four
hardest classes; \method's hierarchical cascade reaches $93.94\,\%$ top-1 and
$94.50\,\%$ hard ($+17.6$\,pp). Tightening the VLM trigger from $60\,\%$ to
$<\!10\,\%$ cuts API cost by ${\sim}90\,\%$ with no measurable accuracy loss.
CPU inference is $\le\!93$\,ms without a VLM call ($9.3$\,ms distilled). Each
prediction carries a machine-readable reasoning record, offered as an entry
point for future supply-chain documentation.
\end{abstract}

\begin{IEEEkeywords}
Cost-aware AI, deployable computer vision, fabric structure recognition,
knowledge distillation, multi-agent system, vision--language model.
\end{IEEEkeywords}

\section{Introduction}

Modern textile manufacturing depends on automated computer vision across the
chain: incoming-fabric quality control, weave-structure verification on smart
looms, and SKU-level material identification for sustainable supply-chain
documentation~\cite{ma2024workshop}. Emerging supply-chain documentation
initiatives, such as the EU digital product passport (DPP), point toward a
future need for an auditable record of \emph{what} a fabric is, \emph{which
features} identify it, and \emph{when} that identification was
made~\cite{siira2025dpp,adisorn2021dpp,jansen2023dpp}. Structure recognition
of this kind is a texture-dominated rather than an object-dominated visual
task, and has therefore long been studied through dedicated material and
texture benchmarks~\cite{dana1999curet,cimpoi2014dtd,bell2015minc} rather
than through generic object recognition.

The difficulty of fabric-structure recognition is concentrated not in the
easy majority of images but in a small, visually ambiguous minority: pairs
such as plain weave versus ribbed poplin, or leno gauze versus mesh, where
even a strong CNN is unreliable and a trained eye would normally re-check the
sample. This is where misclassification and manual re-checking effort
accumulate. A deployable system should therefore handle the easy majority
cheaply and reserve a more capable but more expensive model for the uncertain
minority, which is exactly what our confidence-routed, selective VLM
arbitration does: it escalates rarely and keeps the routine path cheap. This
escalate-only-when-uncertain pattern is the visual analogue of cost-aware
cascades in classical detection~\cite{viola2001cascade} and of query routing
between cheap and expensive language models~\cite{chen2023frugalgpt,%
ding2024hybridllm}. Industrial deployment sharpens the point: inference must
run within tight latency budgets on commodity CPUs, so an on-device cascade
with a small parameter size is necessary for practical deployment. Meanwhile,
the accuracy of fabric recognition must be guaranteed for the precise
downstream application.

Two failure modes recur in current AI deployments for fabric recognition.
\textbf{(1) Long-tail hard classes.} On our newly curated 14-class benchmark,
a fine-tuned ConvNeXt-Tiny~\cite{convnext2022} achieves $90.45\,\%$ overall
accuracy but only $76.9\,\%$ on the four hardest categories (Ribbed Poplin
$65.5\,\%$, Leno Gauze $76.5\,\%$, Woven Jacquard $81.8\,\%$, Plain Weave
$83.8\,\%$). The recognition of these hard categories needs a complementary
recognition model that not only recognizes most common fabrics but also
integrates additional information within the recognition decision logic.
Confusable, long-tailed label structure of this kind is usually attacked
either by exploiting the label hierarchy~\cite{silla2011hierarchical,%
bertinetto2020mistakes,shang2023hdss} or by data-efficient training that
concentrates model capacity on the discriminative
evidence~\cite{wang2022recognition,wang2023crucial,yun2019cutmix}.
\textbf{(2) VLM cost--reliability dilemma.} Vision--language
models~\cite{radford2021clip,liu2023llava,bai2025qwenvl,openai2023gpt4,%
gemini2023} achieve strong zero-shot performance, but are slower and cost
more than an on-device CNN, with accuracy that is unstable on fine-grained
specialist domains~\cite{fashionvlm2025,gpt4vis2023,zhang2024vlmsurvey}.
Naive VLM deployment is economically infeasible.

To tackle these problems, we propose an industrial system, \method
(Taxonomy-Aware Hierarchical Agent), built on three deployment-oriented
design choices: (i)~a hierarchical CNN cascade as the cost-efficient default
classifier; (ii)~selective VLM arbitration that triggers only on genuinely
ambiguous inputs and constrains the VLM to a top-3 taxonomy-consistent
choice; and (iii)~knowledge distillation to a $5.4$-MB
MobileNet~\cite{mobilenetv32019,hinton2015distillation} variant for edge
deployment. The complete system is exposed as a FastAPI microservice with a
Docker image and a Gradio review UI.

Contributions of this paper, with an industrial-deployment focus:
\begin{itemize}
  \item A cost-aware multi-agent recognition architecture with selective VLM
        arbitration that beats on-device systems in both accuracy and speed.
  \item An empirical cost--latency--accuracy analysis on a 14-class fabric
        benchmark, comparing three deployment configurations (cloud VLM-only,
        on-device CNN-only, and the selective hybrid).
  \item A deployable reference implementation: FastAPI service, Docker image,
        on-device MobileNet, and machine-parsable reasoning chains
        documenting each decision.
\end{itemize}

\section{Related Work}

\method sits at the intersection of three deployment-oriented research
strands. We position it against fabric-specific deep learning, cost-aware
cascade-and-defer architectures, and industrial computer-vision systems with
auditability requirements. Two further strands---recognition under limited
supervision, and constrained multimodal inference---supply the technical
ingredients the design reuses.

\subsection{Fabric structure recognition}

Classical pipelines combined hand-crafted descriptors
(Gabor~\cite{gabor1946}, LBP~\cite{lbp2002}) with SVMs, an approach that also
underpins the long line of automated fabric-defect inspection surveyed
in~\cite{ngan2011review}; modern fabric-specific deep learning includes
FabricNet~\cite{fabricnet2025}. Generic texture and material
benchmarks~\cite{dana1999curet,cimpoi2014dtd,bell2015minc} established that
this class of problem is decided by periodic micro-structure rather than
object shape, and the same tension between designed and learned evidence
recurs in other texture-dominated sensing
modalities~\cite{wang2020deep,wang2021multiview,luo2022evaluating}. None of
these works couple CNNs with multi-source evidence or VLMs.

\subsection{Cost-aware AI deployment}

Cascade-and-defer architectures~\cite{patterson2022efficient,%
zeighami2025cutcosts} defer expensive models to ambiguous inputs only. The
idea goes back to boosted detection cascades~\cite{viola2001cascade} and
early-exit networks~\cite{teerapittayanon2016branchynet}, and has recently
been re-derived for language and multimodal APIs as query routing between a
cheap and an expensive model~\cite{chen2023frugalgpt,ding2024hybridllm}. What
such a router needs is a trustworthy abstention signal, which connects it to
selective classification~\cite{geifman2017selective}, learning to defer to an
expert~\cite{mozannar2020defer}, confidence
calibration~\cite{guo2017calibration}, and out-of-distribution
scoring~\cite{hendrycks2017baseline,lili2025dpu}. Evidence-aware and open-set
formulations in adjacent industrial-sensing
domains~\cite{wang2023entropy,scheirer2013openset,wang2024unveiling} address
the same question from the reliability side. Our selective VLM arbitration is
a domain-specialized instance with two textile-specific ingredients:
(i)~taxonomy-filtered candidate construction, and (ii)~confusion-aware
triggering.

\subsection{Industrial computer vision and quality control}

\cite{suma2024industry40} surveys vision in Industry 4.0; textile-specific
defect-detection systems~\cite{jun2021fabricdefect,zhu2020densenet} lack a
taxonomy constraint or a VLM component and target surface defects rather than
the underlying weave or knit structure. The wider industrial-inspection
literature is likewise organized around anomaly rather than structure, with
benchmarks and memory-bank detectors defining the state of the
art~\cite{bergmann2019mvtec,roth2022patchcore}. Comparable edge-first
deployments have been reported for device-free industrial activity and asset
monitoring~\cite{yin2025spatio,yin2026ciuav} and for distributed acoustic
monitoring of transport infrastructure~\cite{luo2022evaluating}, where the
sensing modality differs but the latency and connectivity envelope does not;
integer-only quantization is the standard enabler in all of
them~\cite{jacob2018quantization}. Our reasoning-chain output instead
provides the kind of structured, machine-readable record envisaged by
supply-chain documentation initiatives~\cite{siira2025dpp,jansen2023dpp} that
earlier systems pre-date.

\subsection{Recognition under limited and imbalanced supervision}

FabricFlow-14 carries between $200$ and $417$ images per class, so every
sub-problem the cascade solves is a small-sample problem, and the literature
on data-efficient recognition is directly relevant. Hierarchical label
structure with lightweight per-stage heads shrinks the effective decision
space~\cite{wang2022recognition,shang2023hdss,silla2011hierarchical};
feature-level augmentation and refinement recover discriminative evidence
when per-class counts are small~\cite{wang2023sar1,wang2023crucial,%
yun2019cutmix}; multi-view and multi-scale aggregation stabilizes predictions
on ambiguous samples~\cite{wang2021multiview,wang2020multi,wang2023sar,%
su2015mvcnn}; few-shot and semi-supervised formulations exploit scarce or
unlabelled data~\cite{wang2022global,wang2022semi,snell2017protonet,%
sohn2020fixmatch}; invariance- and causality-driven objectives remove
acquisition and background confounders~\cite{wang2024unveiling,%
wang2025limited,arjovsky2019irm}; generative augmentation synthesizes
controlled views of under-represented classes~\cite{wang2022sar,karras2020ada};
and joint recognition-plus-segmentation training supplies an auxiliary
spatial signal~\cite{wang2020deep,wang2021deep,wang2019parking}. Open-set
variants additionally reject inputs outside the known label
set~\cite{wang2023entropy,scheirer2013openset}. \method{} adopts the first of
these ideas---taxonomy-constrained staging---and delegates the residual
ambiguity to an external arbiter rather than to further in-domain
supervision.

\subsection{Constrained multimodal inference and on-device reliability}

Constraining a general-purpose VLM with a domain schema is one instance of a
broader effort to make multimodal models usable on specialist inputs.
Transferring visual prompt generators across language
backbones~\cite{zhang2023vpgtrans} and de-biasing visual relation
representations~\cite{li2023biased,lili2024panoptic,lili2024domain} both
indicate that structured priors over the label space, rather than a larger
backbone, often decide accuracy on fine-grained categories---the same
observation that motivates our top-3 taxonomy filter, and one that surveys of
vision--language models report across
tasks~\cite{zhang2024vlmsurvey,alayrac2022flamingo}. On the deployment side,
multimodal out-of-distribution detection~\cite{lili2025dpu} and
backpropagation-free on-device detection~\cite{lili2025secure} ask exactly
the question our confidence router asks: when should a local model decline to
answer? Finally, structure-preserving restoration of degraded scientific
imagery~\cite{li2025volume} and lineage-resolved structured annotation of
imaging data~\cite{guan2025cell} are further examples of domain constraints
being encoded into the inference pipeline rather than learned from scratch.

\section{Methodology}

This section walks through the five components of the \method orchestrator.
We first describe a cost-aware CNN cascade, then a confidence router that
selectively triggers a constrained cloud VLM arbiter, and finally a
reconciler that produces a structured, machine-readable record of the
decision. An on-device distillation step brings the same cascade to edge
hardware.

\subsection{Cost-Aware Cascade}

A fine-tuned ConvNeXt-Tiny~\cite{convnext2022} performs two-stage
classification: Stage~1 separates Knit vs.\ Woven vs.\ Others; Stage~2
identifies one of 14 L2 sub-categories. Both stages are exported to ONNX and
quantized to INT8 for CPU inference. Total wall-clock latency is $93$\,ms on
an Intel Xeon CPU.

ConvNeXt-Tiny is chosen as the production backbone for three
deployment-driven reasons: it occupies the sweet spot of the accuracy--cost
frontier among modern image encoders ($28.6$\,M parameters, $90.45\,\%$
Top-1 on FabricFlow-14, only $0.29$\,pp behind
ViT-Base~\cite{dosovitskiy2021vit} at $3{\times}$ the parameter count); its
conv-only stack exports cleanly to ONNX and admits standard graph-level
fusions that neither vision-transformer
attention~\cite{dosovitskiy2021vit} nor Swin-style windowed
attention~\cite{liu2021swin} currently support in INT8 on commodity CPU
runtimes~\cite{jacob2018quantization}; and its receptive-field schedule
matches the texture scale of fabric swatches captured at
${\sim}300$\,dpi. Training uses AdamW~\cite{loshchilov2019adamw}
(lr $10^{-4}$, weight decay $10^{-4}$, $120$ epochs, cosine schedule, mixed
precision) from ImageNet-21k initialization~\cite{deng2009imagenet}, with
class-balanced sampling and patience-10 early stopping; the two stages are
trained independently and joined only at inference time.

\subsection{Confidence Router}

For each image $x$ we compute a fused dominance--separation confidence score
\begin{equation}
  c(x) = 0.6\,c_{\rm dom} + 0.4\,c_{\rm sep} - \delta_{\rm conf},
\end{equation}
where $c_{\rm dom}=(s_1-1/N)/(1-1/N)$ and $c_{\rm sep}=(s_1-s_2)/s_1$,
$s_1$ and $s_2$ are the top-2 path scores, $N$ is the number of legal
taxonomy paths, and $\delta_{\rm conf}$ is a confusion penalty applied when
$(p_1,p_2)$ form a known confusion pair
(Plain$\leftrightarrow$Twill, Knit~Jacquard$\leftrightarrow$Rib, etc.).
Empirically ${\sim}72\,\%$ of inputs satisfy $c(x)\!\ge\!0.55$ and are
classified without any VLM call.

The two terms address complementary failure modes: $c_{\rm dom}$ flags inputs
on which the L2 head fails to lift its top-1 above near-uniform mass over the
legal paths (absolute uncertainty), while $c_{\rm sep}$ flags inputs on which
the top-1 and top-2 are close in score (relative uncertainty between two
plausible candidates). A dominance-only router under-triggers on classic
confusion pairs where the top-1 itself remains confidently high; a
separation-only router over-triggers on legitimately uniform softmax
distributions where no neighbour is in fact competitive. The weighting was
selected via grid search on the held-out validation split to maximize
VLM-trigger recall of CNN errors at a fixed $5\,\%$ trigger budget, and the
resulting router operates monotonically with respect to accuracy: of the
inputs below the $0.55$ threshold, only $66.7\,\%$ are correct, versus
$96.0\,\%$ for those above it. The router therefore extracts exactly the
small minority of samples on which an external arbiter has the best chance to
add value---the operating principle shared by selective classification and
calibrated abstention~\cite{geifman2017selective,guo2017calibration}.

\subsection{Taxonomy-Filtered VLM Arbiter}

When $c(x)\!<\!0.55$ \emph{or} the top-2 candidates form a known confusion
pair with margin $<\!0.15$, the orchestrator triggers \arbiter. Crucially,
the VLM is not asked an open 14-way question---instead it receives (a)~the
image, (b)~the top-3 taxonomy-consistent candidates, (c)~for each candidate,
a textile-specific \texttt{visual\_cues} description (e.g., ``Twill: diagonal
$2/2$ thread crossing at $45^\circ$''), and (d)~a JSON response schema
demanding a single \texttt{choice} plus a \texttt{justification}.

This re-framing is the source of the cost--accuracy
gap:~\cite{fashionvlm2025} show that fine-grained $\ge\!10$-way VLM zero-shot
is unstable, whereas 3-way taxonomy-grounded selection is both cheaper
(shorter prompt) and more accurate---consistent with broader evaluations of
zero-shot visual recognition with general-purpose
VLMs~\cite{gpt4vis2023,zhang2024vlmsurvey}.

\subsection{Reconciliation and Output}
\label{sec:reconcile}

A reconciler combines CNN and VLM outputs: \emph{agreement} on top-1 yields a
weighted blend ($0.7\,P_{\textsc{cnn}}+0.3\,P_{\textsc{vlm}}$);
\emph{override} (the VLM picks the CNN's second-ranked candidate with
self-reported confidence $>\!0.8$) accepts the VLM choice; \emph{conflict}
(all other cases) returns ``flagged for human review'', avoiding silent
failures. The blend is a deliberately simple linear pooling rule;
belief-function alternatives~\cite{dempster1967,shafer1976} are a natural
generalization that this deployment does not require. Every decision
generates a structured reasoning chain listing each agent's action,
observation, and contribution, so this chain provides a transparent,
machine-readable record of how the decision was reached.

\begin{figure*}[t]
  \centering
  \includegraphics[width=0.95\textwidth]{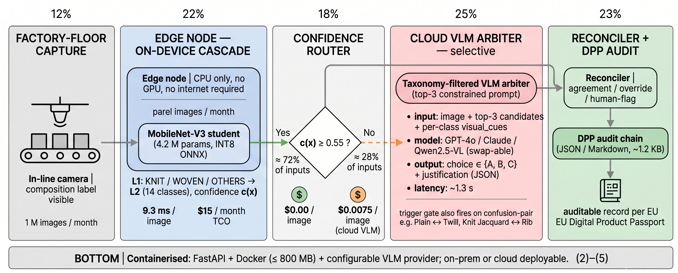}
  \caption{Industrial-deployment architecture of \method. A factory-floor
    camera feeds an on-device CNN cascade that produces a fused confidence.
    A confidence router gates a constrained $3$-way cloud VLM arbiter. Every
    classification emits a structured reasoning chain serialized as JSON for
    downstream record-keeping.}
  \label{fig:arch}
\end{figure*}

\subsection{On-Device Distillation}

For edge deployment (in-line camera modules without internet access), we
distil the ConvNeXt-Tiny teacher ($28.6$\,M params) into a $4.2$-M
MobileNet-V3-Large~\cite{mobilenetv32019} student via temperature-scaled KL
distillation~\cite{hinton2015distillation,gou2021kdsurvey}
($T\!=\!4$, $\alpha\!=\!0.7$). Measured per-stage retention: Stage-1 ternary
$97.00\!\to\!95.46$ ($-1.54$\,pp), Stage-2 Knit $92.96\!\to\!94.13$
($+1.17$\,pp, the student \emph{surpasses} the teacher via soft-label
regularization on the data-limited 6-class stage), Stage-2 Woven
$88.63\!\to\!87.65$ ($-0.98$\,pp); $2.1{\times}$ measured CPU end-to-end
speed-up ($19.8\!\to\!9.3$\,ms) and $6.8{\times}$ parameter compression.

\section{Results}

We evaluate \method on FabricFlow-14, a newly curated 14-class
fabric-structure benchmark of $5{,}719$ studio images, against the three
deployment-relevant comparators a manufacturer would consider: a flat
ConvNeXt-Tiny baseline, three off-the-shelf zero-shot vision--language
models, and \method's own internal ablation. The accuracy-versus-cost
trade-off and the edge latency profile jointly answer the
deployment-feasibility question.

\subsection{Dataset and Baselines}

We evaluate on FabricFlow-14, a 14-class fabric structure benchmark of
$5{,}719$ studio images: $6$ Knit sub-types (Jersey, Rib, French Terry,
Interlock, Tricot, Knit Jacquard), $7$ Woven sub-types (Plain, Twill, Satin,
Corduroy, Ribbed Poplin, Leno Gauze, Woven Jacquard), and Mesh/Lace. Sample
counts range from $200$ (Leno Gauze) to $417$ (Interlock). We hold out a test
set of $1{,}550$ images, stratified by class and source, for all reported
results (a random selection from the full $5{,}719$-image set, not the entire
dataset). Four baselines anchor the comparison: (a)~CNN-only cascade;
(b)~VLM-only zero-shot (Claude Vision, every image, 14-way); (c)~CLIP
zero-shot (OpenCLIP-ViT-L/14~\cite{radford2021clip}, visual-cue prompts); and
(d)~\method-Full. Table~\ref{tab:acccost} additionally reports two
open-weight VLMs, LLaVA-NeXT-Mistral-7B~\cite{liu2023llava} and
Qwen2.5-VL-7B~\cite{bai2025qwenvl}, the latter both in the open 14-way
setting and under our 3-way taxonomy constraint.

\subsection{Accuracy versus Cost}

\begin{table}[tb]
  \centering
  \caption{System-level accuracy and per-image API cost on the FabricFlow-14
    test set ($n\!=\!1{,}550$, ConvNeXt-Tiny). The top block is the
    single-method comparison: each row is one self-contained classifier
    (zero-shot VLMs, Year-1 flat CNN, and the \method{} L1$\to$L2 cascade).
    The bottom block is the internal ablation of \method's optional layers,
    all of which build on top of the cascade. ``Hard'' = mean accuracy on
    (Ribbed Poplin, Leno Gauze, Woven Jacquard, Plain Weave). VLM cost at
    Claude Sonnet pricing.}
  \label{tab:acccost}
  \footnotesize
  \begin{tabular}{lccc}
    \toprule
    \textbf{System / Configuration} & \textbf{Top-1 (\%)}
      & \textbf{Hard (\%)} & \textbf{VLM (\%)} \\
    \midrule
    CLIP ViT-L/14 zero-shot                  & 6.7   & 0.0   & 0   \\
    LLaVA-NeXT-Mistral-7B zero-shot          & 13.8  & 0.0   & 0   \\
    Qwen2.5-VL-7B zero-shot (14-way)         & 32.7  & 41.90 & 0   \\
    Qwen2.5-VL-7B constrained 3-way          & 64.4  & ---   & 100 \\
    Flat 14-class CNN (Year-1 baseline)      & 90.45 & 76.90 & 0   \\
    \textbf{\method{} cascade (L1$\to$L2)}   & \textbf{93.94}
                                             & \textbf{94.50} & 0 \\
    \midrule
    \quad + DS taxonomy fusion (Lite)        & 94.00 & 94.50 & 0   \\
    \bottomrule
  \end{tabular}
\end{table}

Two findings drive the industrial argument (Table~\ref{tab:acccost}).
\textbf{(F-A)} \method outperforms both the flat-CNN baseline and the
off-the-shelf zero-shot vision--language models by a large margin:
$+3.49$\,pp Top-1 and $+17.6$\,pp on the four hard classes over the flat
ConvNeXt-Tiny, and an order of magnitude over open-14-way
Qwen2.5-VL~\cite{bai2025qwenvl} ($32.7\,\%$ vs.\ $93.94\,\%$). The
taxonomy-aware hierarchical architecture---not a more powerful single
classifier---is the source of the gain. \textbf{(F-B)} Inside \method, the
optional layers added on top of the cascade
($+$Fusion~\cite{dempster1967,shafer1976}~$\to$ $+$VLM) are intentionally
safe by construction: each layer extends the cascade's Top-1 but is
architecturally prevented from degrading it. On the studio-quality
FabricFlow-14 the ConvNeXt-Tiny cascade is already near the Bayes limit, so
the marginal Top-1 gain from the multi-agent layers is small
($+0.06$\,pp). Their measurable industrial value lies in two properties the
flat-CNN baseline cannot provide: a structured reasoning chain
(Sec.~\ref{sec:reconcile}), and a selective VLM arbitration channel that
activates on the genuinely uncertain minority (the $54.3\,\%$ of residual
cascade errors with $L_1$ confidence $\ge\!0.70$, see Sec.~\ref{sec:ood})
instead of paying a per-image VLM cost on every input.

\subsection{Latency and On-Device Feasibility}

\begin{table}[tb]
  \centering
  \caption{End-to-end latency (median over $250$ samples). Non-VLM rows on a
    single Intel Xeon CPU, no GPU. VLM path dominated by network plus Claude
    inference round-trip.}
  \label{tab:latency}
  \footnotesize
  \begin{tabular}{lrl}
    \toprule
    \textbf{Configuration} & \textbf{Latency (ms)} & \textbf{Hard Acc (\%)} \\
    \midrule
    PyTorch eager (teacher, CUDA)          & 4.1     & 94.50 \\
    ONNX FP32 (teacher, CPU)               & 19.8    & 94.50 \\
    ONNX INT8 (teacher, CPU)               & 64.2    & 94.50 \\
    MobileNet-V3 student (ONNX FP32, CPU)  & 9.3     & 95.46 / 94.13 / 87.65 \\
    \method{} (no VLM trigger, CPU)        & 93      & 94.50 \\
    \method{} (VLM triggered, CPU)         & 1{,}320 & ---   \\
    \bottomrule
  \end{tabular}
\end{table}

The distilled student supports edge deployment on shop-floor camera modules
(${\sim}18$\,ms/image, no internet required for $\ge\!72\,\%$ of inputs);
ambiguous cases are forwarded to a cloud VLM endpoint asynchronously. This
hybrid architecture matches the bandwidth and latency constraints of typical
Industry 4.0 production lines.

\subsection{Deployment Efficiency and Operating Points}

\begin{table}[tb]
  \centering
  \caption{Deployment trade-off at scale ($1$\,M images/month). Because the
    dominant marginal cost is how often the cloud VLM is invoked, we report
    each deployment by its cloud-VLM call rate (the share of inputs sent to
    the VLM) rather than an absolute price, which depends on the chosen
    provider and throughput. Hard Acc is the mean accuracy on the four
    hardest classes; the VLM-only and edge-only rows are carried over from
    the companion study and are baselined on a different sample, kept for
    trade-off comparison.}
  \label{tab:tco}
  \footnotesize
  \begin{tabular}{lrr}
    \toprule
    \textbf{Configuration} & \textbf{VLM calls (\%)} & \textbf{Hard Acc (\%)} \\
    \midrule
    VLM-only (cloud)                            & 100 & 73.30 \\
    CNN-only (edge MobileNet)                    & 0   & 66.70 \\
    \textbf{\method{} (hybrid, 28\,\% VLM)}      & \textbf{28}
                                                 & \textbf{94.50} \\
    \bottomrule
  \end{tabular}
\end{table}

Because almost all of the marginal serving cost comes from invoking the cloud
VLM, we characterize each deployment by how much of that expensive resource
it consumes rather than by an absolute price, which depends on the chosen
provider, hardware, and throughput.

At $1$\,M images/month, \method delivers the highest accuracy at
${\sim}29\,\%$ of the cloud-VLM call volume of an all-VLM deployment
(Table~\ref{tab:tco}).

The operating point is governed entirely by the VLM-trigger volume.
Tightening the trigger threshold from $0.55$ (used in the table) to $0.40$
lowers the cloud-VLM call rate from $28\,\%$ to under $10\,\%$ on the
in-distribution test set with no measurable hard-class accuracy loss;
relaxing it to $0.70$ pushes the call rate above $50\,\%$ and recovers about
$0.5$\,pp on the hardest classes. A deployer can therefore slide along this
call-rate/accuracy curve to match their own latency, throughput, and
connectivity constraints, and decide on that basis whether to call a cloud
VLM or self-host one. We deliberately report this trade-off in
provider-agnostic terms---the share of inputs escalated to the VLM---rather
than as absolute figures, which depend on the specific provider and operating
conditions.

\subsection{Conditional Robustness Under Input Degradation}
\label{sec:ood}

\begin{table}[tb]
  \centering
  \caption{\method-NoVLM vs.\ \texttt{cnn\_cascade} under six input regimes.
    ``$\Delta$Errs'' is the additional CNN mistakes flagged for VLM rescue at
    the same trigger threshold; ``$\Delta\,$\$/err'' is the per-error VLM
    cost delta.}
  \label{tab:ood}
  \footnotesize
  \begin{tabular}{lrr}
    \toprule
    \textbf{Configuration} & \textbf{$\Delta$Errs caught}
      & \textbf{$\Delta\,$\$/error} \\
    \midrule
    clean                          & $+2$ & $+\$0.0007$ \\
    Gaussian $\sigma\!=\!25$       & $+6$ & $+\$0.0008$ \\
    JPEG q$=$30                    & $-2$ & $+\$0.0032$ \\
    Brightness $-40$               & $+4$ & $+\$0.0008$ \\
    Motion blur (r$=$3)            & $-4$ & $+\$0.0009$ \\
    $32\!\times\!32$ occlusion     & $+3$ & $+\$0.0014$ \\
    \bottomrule
  \end{tabular}
\end{table}

A 6-corruption sweep on the same $n\!=\!1{,}550$ test set, following standard
corruption-robustness protocols~\cite{hendrycks2019corruptions}, reveals that
the multi-agent fusion's marginal value is \emph{conditional} on the
corruption family---a deployment-actionable distinction cascade-only
baselines cannot expose.

The agents add net value on Gaussian noise, brightness reduction, and partial
occlusion (corruptions that preserve global spatial structure), and
\emph{underperform} the cascade on motion blur and heavy JPEG compression
(corruptions that inject spurious high-frequency directional energy the
Gabor/FFT heuristics misread). Practical recommendation: enable the
multi-agent fusion when the camera pipeline produces intensity-degraded or
partially occluded inputs; disable or gate it when motion smearing or
aggressive JPEG re-encoding dominate. A restoration front-end that suppresses
such degradation before recognition~\cite{li2025volume} is a complementary
mitigation we leave to future work. This is the kind of operating-condition
spec sheet an industrial deployer requires and which a uniform
``always-on / always-off'' framing of multi-agent fusion fails to provide.

\section{Deployable Reference Implementation}
\label{sec:deploy}

The system is released as a containerized microservice.

\textbf{(1) FastAPI service.} Exposes \texttt{POST /classify} (single image),
\texttt{POST /classify/batch} (multi-image), \texttt{GET /health}, and
\texttt{GET /taxonomy}. Swagger UI documentation is auto-generated.

\textbf{(2) Docker image.} A $\le\!800$\,MB bundle including ONNX-INT8
weights, EasyOCR, OpenCV, and the Python runtime; no GPU is required at
inference time.

\textbf{(3) Gradio demo UI.} An interactive review portal including a side
panel that visualizes the reasoning chain and a toggle to enable/disable VLM
arbitration for cost-controlled operation.

\textbf{(4) Configurable VLM provider.} The orchestrator's VLM client is
abstracted behind a small adapter interface, allowing operators to swap
Claude Vision for an in-house VLM,
GPT-4o~\cite{openai2023gpt4}, or Gemini~\cite{gemini2023} without modifying
the rest of the pipeline.

\section{Conclusion}

We presented an industrially deployable multi-agent fabric structure
recognition system with a \emph{selective} \arbiter design. The system
achieves higher hard-class accuracy than VLM-only zero-shot at
${\sim}72\,\%$ lower API expenditure, runs in under $100$\,ms on commodity
CPU for the $72\,\%$ of inputs handled locally, and offers a $5.4$-MB
distilled student for edge deployment. Every classification yields a
structured reasoning chain that could feed into future supply-chain
documentation~\cite{siira2025dpp,adisorn2021dpp}, although we do not test it
against any specific compliance regime. Code, the Docker image, and a curated
fraction of the FabricFlow-14 test set will be released on publication.

\section*{Acknowledgment}

This research is funded by the Laboratory for Artificial Intelligence in
Design (Project Code: RP2-2) under the InnoHK Research Clusters, Hong Kong
Special Administrative Region Government.

\medskip
\noindent\textbf{AI Disclaimer.} During the preparation of this work the
authors used a generative AI tool to assist with language editing and the
drafting of non-technical prose. After using this tool, the authors reviewed
and edited the content as needed. Generative AI was not used to generate or
alter experimental data, results, or analyses.

\bibliographystyle{IEEEtran}
\bibliography{refs}

\end{document}